\documentclass[fleqn,usenatbib]{mnras}

\usepackage{amsmath}
\usepackage{amssymb}
\usepackage{epsfig}
\usepackage{ifthen}
\usepackage{latexsym}
\usepackage{rotating}
\usepackage{times,epsf}
\usepackage{txfonts}
\usepackage{varioref}
\usepackage{verbatim}
\usepackage{url}
\usepackage{color}
\usepackage{array}
\usepackage[dvipsnames]{xcolor}
\usepackage[T1]{fontenc}
\usepackage{multirow}
\usepackage{graphicx}

\newcolumntype{M}{>{$\vcenter\bgroup\hbox\bgroup}c<{\egroup\egroup$}}

\newcommand{\be}{\begin{equation}}
\newcommand{\ee}{\end{equation}}
\newcommand{\bea}{\begin{eqnarray}}
\newcommand{\eea}{\end{eqnarray}}

\title[Radial oscillations of levitating atmospheres]{Radial oscillations of a
radiation-supported levitating shell in Eddington luminosity neutron
stars}

\author[D. Abarca, W. Klu{\'z}niak]{
David Abarca$^1$, 
W\l odek Klu{\'z}niak$^1$
\thanks{
E-mail: dabarca@camk.edu.pl (DA) ;
wlodek@camk.edu.pl (WK)}\\
$^1$ Nicolaus Copernicus Astronomical Center, Bartycka 18, Warsaw
00-716, Poland}

\begin{document}
\maketitle

\label{firstpage}

\begin{abstract}
In general relativity, it has been shown that radiation-supported atmospheres exist outside 
the surface of a radiating spherical body close to a radius where the 
gravitational and radiative forces 
balance each other. 
We calculate the frequency of oscillation of the incompressible radial mode
of such a thin atmospheric shell and show that in the optically thin case, this particular mode 
is overdamped by radiation drag. 

\end{abstract}


\section{Introduction}
Neutron stars have been shown to erupt in thermonuclear (Type I) X-ray bursts. 
In addition,
pulsations of an Ultraluminous X-ray Source (ULX) have been explained by accretion 
onto a neutron star \citep{bachetti+14}. At such large ($\sim 100$ times Eddington) luminosities, it is easy to 
imagine that neutron star systems may produce luminosities above the 
Eddington limit. 

In Newtonian physics, the radiative force is proportional to 
the flux, which falls off as $1/r^2$ for a spherically symmetric source. If the radiative force
exceeds the gravitational force at one radius, then it will exceed it at all radii. 
In general relativity the radiative force can be shown to increase faster than the 
gravitational force at smaller radii \citep{phinney+87,abramowicz}. 
It turns out that for a given luminosity there
exists a radius where the gravitational and radiative forces are equal, forming 
an imaginary, spherically symmetric surface referred to as the Eddington Capture Sphere (ECS)
\citep{stahl+12, wielgus+12} . 
\cite{wielgus+thin} have shown that it is possible to create an
optically thin atmosphere at this 
radius which levitates above the surface of the star, supported entirely by 
radiation. \cite{wielgus+thick} have extended the analysis to include
optically thick atmospheres as well.  

We are interested if oscillations of these atmospheres can explain the
still unresolved problem of the source of Quasi-periodic Oscillations (QPOs), 
the transient peaks observed in the power spectrum of highly compact sources \citep{remillard+06},
including neutron stars \citep{vanderklis+06}. 
Specifically in X-ray bursting neutron stars, there have been observations of hectoHertz QPOs; \cite{strohmayer+99} report on several QPOs from 
X-ray bursts from low mass X-ray binaries (LMXB) all with frequencies between 300-600 Hz. 
Moreover, most sources show an increase of frequency with time during the decay phase
\citep{strohmayer+99,strohmayer+2006}.
 
We investigate the possibility of oscillations of atmospheres around the ECS
which could possibly provide an explanation for X-ray burst hectoHertz 
QPOs. We begin by finding an incompressible radial oscillation mode and 
calculating its oscillation frequency. We then compute the effects of radiation 
drag and discuss the viability of such a mode to produce a QPO.

\section{Eigenmode of an Incompressible Thin Shell}
\label{thinshell}
First we demonstrate the equations that
are used to construct the atmospheres from
\cite{wielgus+thin}. This is essentially the 
relativistic equation for hydrostatic 
equilibrium for an optically thin fluid
subject to the radiative force from a 
spherical source. We also explicitly include
the derivation of the ECS from \cite{stahl+12},
because the equations involved are also used to 
calculate the frequency of oscillations of a thin
shell about the ECS, as well as the contribution
of radiation drag. 
 
 \subsection{Relativistic Hydrostatic Equilibrium}
For this relativistic calculation
we use the Schwarzschild metric, 
\begin{equation}
ds^2 = - B(r) dt + B(r)^{-1} dr + r^2 d\Omega^2,
\end{equation}
where $B(r) = 1-2r_g/r$ for $r_g = GM/c^2$. We find it convenient to use units where
$G=c=1$. 

Let us consider the equation for the conservation of stress-energy given by
\begin{equation}
\nabla_\mu T^{\mu\nu}=0,
\end{equation}
where $T^{\mu\nu}$ is the stress-energy tensor for a perfect fluid given by, 
$T^{\mu\nu} = (p+\rho+u)u^\mu u^\nu + p g^{\mu \nu}$, for pressure, $p$, rest mass density,
$\rho$, and internal energy density, $u$. We can project the conservation equation 
onto the space orthogonal to the four-velocity using the projection tensor, 
$h^{\mu\nu}=u^\mu u^\nu + g^{\mu\nu}$, to get the relativistic Euler equation, 
\begin{equation}
\label{eqn:euler}
u^\mu \nabla_\mu u^\nu + \dfrac{h^{\mu\nu}  \nabla_\mu p}{p+\rho+u}=f^\mu ,
\end{equation}
where we have also added a four-force, $f^\mu$, which corresponds to the radiation 
force due to Thomson scattering
 with cross section, $\sigma$, in an optically thin fluid around a luminous star, which can be written
 in terms of the radiative flux, $F^\mu$, as 
 \begin{equation}
 f^\mu = \dfrac{\sigma }{m}F^\mu.
 \end{equation}
 Here, $m$ is the proton mass. 

One can construct an atmosphere with its pressure maximum located at, $r_0$, the radius of 
the ECS. These atmospheres obey the equation of hydrostatic equilibrium which can 
be calculated by substituting $u^\mu = u^t(1,0,0,0)$ into the relativistic Euler equation, which gives
\begin{equation}
\label{eqn:hse}
\dfrac{1}{\rho}\dfrac{\partial p}{\partial r}=-\dfrac{M}{1-2M/r}\left(\dfrac{1}{r^2}-\dfrac{f^r}{M}\right).
\end{equation}

\subsection{Derivation of the Equilibrium Surface}
Let us now demonstrate a quick derivation 
of, $r_0$, 
the radius of the ECS.
We start with the expression in parentheses 
in Eq.~\ref{eqn:hse}, which we will eventually
set to zero.

For convenience we name it $\mathcal{F}(r,0) =  1/r^2-f^r/M$, the reason for which will be
explained in the next section.
In terms of the flux, $F^r$ we have 
\begin{equation}
f^r =\dfrac{\sigma}{m}F^r.
\end{equation} 

Expressions for the flux can be found in \cite{stahl+13}. For a stationary 
particle we have
\begin{equation}
F^r= T^{tr} u_t .
\end{equation}

There $T^{tr}$ is the $rt$ component of the radiation 
stress-energy tensor, $T^{\mu\nu}$,
outside a luminous star, first derived in \cite{abramowicz}.
We have
\begin{equation}
T^{rt} = \pi I(r) \sin^2\alpha(r),
\end{equation}
for intensity, $I(r)$, and viewing angle, $\alpha$, defined as 
\begin{equation}
\alpha(r) = \arcsin\dfrac{R}{r}\dfrac{\sqrt{1-2M/r}}{\sqrt{1-2M/R}},
\end{equation}
where $R$ is the radius of the star.

We can write the intensity in terms of the luminosity at infinity, $L_\infty$,
\begin{equation}
I(r) = \dfrac{L_\infty}{4\pi^2R^2}\dfrac{1-2M/R}{(1-2M/r)^2}.
\end{equation}

Putting all of this back into the expression for the acceleration and 
using the usual expression for the Eddington luminosity, $L_\mathrm{Edd} = 4\pi  M m /\sigma$, 
we get
\begin{equation}
f^r= \dfrac{L_\infty}{L_{\text{Edd}}}\dfrac{M}{r^2\sqrt{1-2M/r}} .
\end{equation}
For brevity we define, $\lambda = L_\infty/L_\text{Edd}$. 
We can write down the equation for $\mathcal{F}(r,0)$ for a stationary fluid,
\begin{equation}
\mathcal{F}(r,0) = \dfrac{1}{r^2}\left(1-\dfrac{\lambda}{\sqrt{1-2M/r}}\right).
\end{equation}
\begin{figure}

\includegraphics[width=\columnwidth]{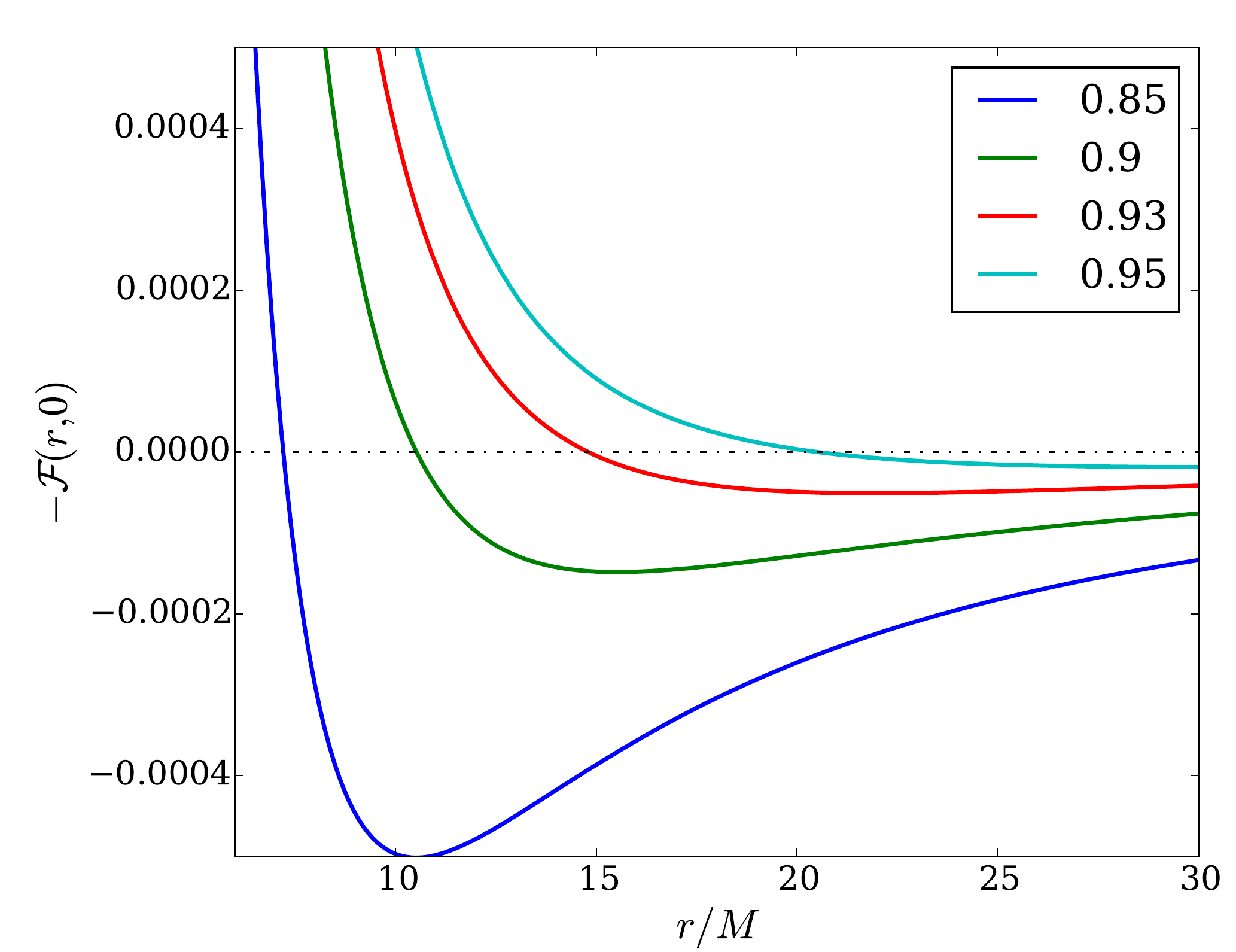}
\caption{Plots of -$\mathcal{F}(r,0)$ for various values of $\lambda$ shown in the legend. 
The location where 
$\mathcal{F}(r,0)=0$ shows the radius of the 
ECS. -$\mathcal{F}(r,0)$ shows the forces
acting on a stationary fluid, and so one can
see how atmospheres can be constructed around
the ECS.
\label{fig:f_of_x}}
\end{figure}

\begin{figure}
\includegraphics[width=\columnwidth]{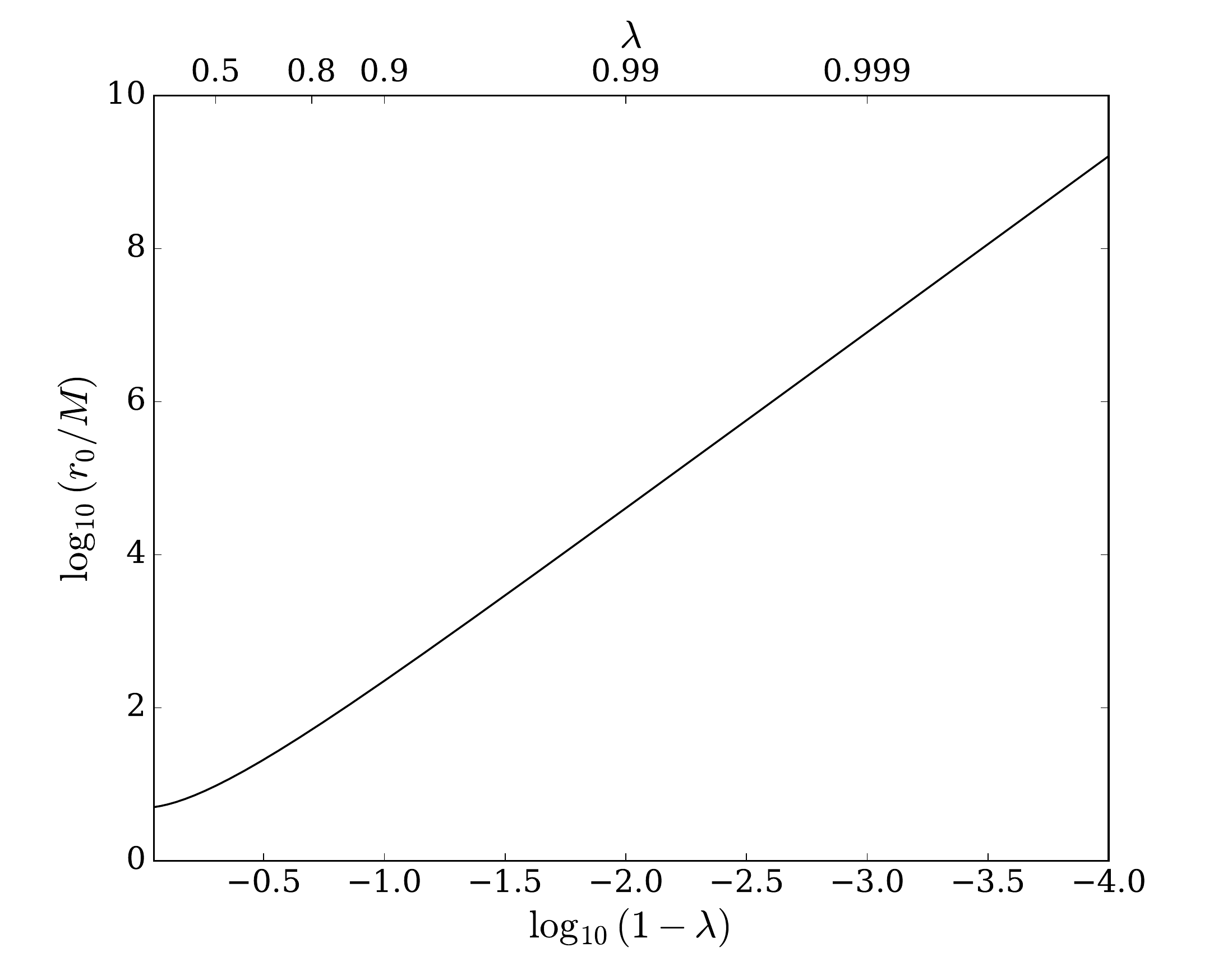}
\caption{We have plotted $r_0$ as a function of $\lambda$. To illustrate
the sensitivity of $r_0$ to $\lambda$, we have plotted against two different axis scales.
The upper is $\lambda$ and the lower shows $\log(1-\lambda)$.\label{fig:x0}}
\end{figure}

We note the location of the ECS to be where $\mathcal{F}(r_0,0)=0$,
this gives
\begin{equation}
r_0 = \dfrac{2M}{1-\lambda^2}.
\end{equation}
 It is important to note that 
$r_0$ is extremely sensitive to $\lambda$ as $\lambda \rightarrow 1$. 
This is further illustrated in Fig.~\ref{fig:x0}.

\subsection{Atmospheric Solution}
The equation for hydrostatic
equilibrium, Eq.~\ref{eqn:hse} was solved
in \cite{wielgus+thin}, who took an
optically thin fluid  
at a sufficiently low temperature such that $\rho \gg p + u$,
and derived atmospheric solutions with their pressure maximum located at radius $r_0$.
They have provided a set of polytropic atmosphere solutions
with polytropic index, $\Gamma$, given by
\begin{equation}
\rho(r) = \rho_0 
\left[
	\ln(1-2M/r)^{-1/2} - \lambda (1-2M/r)^{-1/2}+\lambda
\right]^{\frac{1}{\Gamma -1}}.
\end{equation}
Such an atmosphere is entirely supported by radiation and there are regions between the atmosphere and the surface 
of the star with no gas at all.

\section{Radial eigenmode of an incompressible
thin shell}
Now we turn from hydrostatic equilibrium to the 
time dependent equation, the relativistic Euler
equation, with which we will apply perturbations
to derive the oscillation frequency.
While we are in the relativistic regime of strong gravity, we will make the assumption that our velocities are small
and so non-relativistic. This allows us to write the four velocity, $u^\mu$, to first order in $u^r=dr/d\tau$, or in $v=dr/dt$.
\begin{equation}
u^\mu =(u^t,u^r,0,0)= u^t(1,v,0,0)\approx \dfrac{1}{\sqrt{1-2M/r}}(1,v,0,0).
\end{equation}
This allows us to simplify Eq.~\ref{eqn:euler} to get,
\begin{equation}
\dfrac{d}{d\tau}u^r + 
\dfrac{1}{\rho}\left( g^{rr} \dfrac{\partial p}{\partial r} + u^r \dfrac{dp}{d\tau} \right)
= -M\left(\dfrac{1}{r^2}-\dfrac{f^r}{M}\right) = - M \mathcal{F}(r,u^r),
\end{equation}
We have combined the two non-fluid forces into one expression, $\mathcal{F}$, which is in
general a function of $r$ and $u^r$. We will demonstrate that $\mathcal{F}$ 
can be divided into the sum of
two terms, one of which is only a function of $r$,
the other a function of both $r$ and $u^r$. The
former corresponds to the radiation force, and the
latter to radiation drag, both of which will play
an important role in our analysis. 

We let Eq.~\ref{eqn:hse} serve as the background over which we consider spherically symmetric radial perturbations.
In this work, we will consider an incompressible mode, where the whole atmosphere is transported
by a small radial distance, $\xi$, while preserving 
its pressure and density profiles, $p(r)\rightarrow p_b(r-\xi)$, and $\rho(r) \rightarrow \rho_b(r-\xi)$, where the $b$ index
indicates the background solutions. 
We also  have, $u^r = d\xi/d\tau$.
We will demonstrate
that for a sufficiently thin atmosphere, we will recover a radial eigenmode that oscillates with the same frequency
as a test particle about $r_0$.
This gives us the following two equations to solve, 
\begin{align}
\dfrac{d^2\xi}{d\tau^2} + 
\dfrac{g^{rr}(r)}{\rho_b(r-\xi)}\left(  \dfrac{\partial }{\partial r}p_b(r-\xi) \right)&= - M \mathcal{F}(r,u^r) \label{eqn:perturbed}, \\
\dfrac{g^{rr}(r)}{\rho_b(r)}\dfrac{\partial p_b(r)}{\partial r} &=-M\mathcal{F}(r,0),\label{eqn:simplehse}
\end{align}
where we have been explicit with the $r$ dependence 
for clarity. The system can be simplified to one equation after expanding in terms 
of $\xi$, and substituting Eq.~\ref{eqn:simplehse}
into Eq.~\ref{eqn:perturbed} to get,
\begin{equation}
\dfrac{1}{M}\dfrac{d^2\xi}{d\tau^2}-\mathcal{F}(r,0) + \dfrac{\partial \mathcal{F}(r,0)}{\partial r}\xi
- \mathcal{F}(r,0) \dfrac{\partial \ln g^{rr} (r)}{\partial r}\xi 
= - \mathcal{F}(r,u^r).
\end{equation}

At this point we invoke the thin shell limit. We assume that we have a thin atmosphere concentrated around, 
$r_0$. Also note that, $\mathcal{F}(r_0,0)=0$,
which we derived in the previous section.
This allows us to expand $\mathcal{F}$ around $r=r_0$ and $u^r=0$ to get, 
\begin{equation}
\mathcal{F}(r,u^r)\approx (r-r_0) \left.\dfrac{\partial \mathcal{F}}{\partial r}\right|_{r_0,0} + 
u^r \left.\dfrac{\partial \mathcal{F}}{\partial u^r}\right|_{r_0,0}.
\end{equation} 
Substituting in the expansion and noting that $\xi\cdot (r-r_0)$ is second order in small quantities, we are
left with the following equation, 
\begin{equation}
\label{eqn:dho}
\dfrac{1}{M}\dfrac{d^2\xi}{d\tau^2} +\left.\dfrac{\partial \mathcal{F}}{\partial u^r}\right|_{r_0,0} \dfrac{d\xi}{d\tau}
 +  \left.\dfrac{\partial \mathcal{F}}{\partial r}\right|_{r_0,0} \xi =0.
\end{equation}
This is the equation for a damped harmonic oscillator (assuming the appropriate signs for the derivatives of
$\mathcal{F}$). The equation of motion contains no fluid terms and so we expect this particular incompressible 
mode to oscillate about the equilibrium position with the same frequency as a test particle,
in fact, the trajectories should be identical
in the limit of small perturbations.

\section{Oscillations}
\subsection{Undamped Oscillation Frequency}
Our equation of motion in the thin shell limit is given by Eq.~\ref{eqn:dho}.
In order to calculate the 
frequency of oscillations, first we will 
neglect the damping (second) term in Eq.~\ref{eqn:dho}.

If 
\begin{equation}
\left. \dfrac{\partial \mathcal{F}}{\partial r} \right| _{r_0,0}>0,
\end{equation}
then we can expect the harmonic oscillator solution, 
with angular frequency, 
\begin{equation}
\tilde{\omega}=\sqrt{M\left.\dfrac{\partial \mathcal{F}}{\partial r}\right|_{r_0,0}}.
\end{equation}
We can tell this is true from Fig.~\ref{fig:f_of_x}, but we will calculate it explicitly to find the frequency
that the atmosphere would oscillate at if the radiation drag were negligible. 

Taking the derivative and substituting for the radius of the equilibrium position in terms of $\lambda$ we get
\begin{equation}
\tilde{\omega} = \dfrac{(1-\lambda^2)^2}{4M\lambda}.
\end{equation}

To put our angular frequency in terms of s$^{-1}$ we restore $c$ and $G$ to get
$\omega = c^3\tilde\omega/G$.

\begin{figure}

\includegraphics[width=\columnwidth]{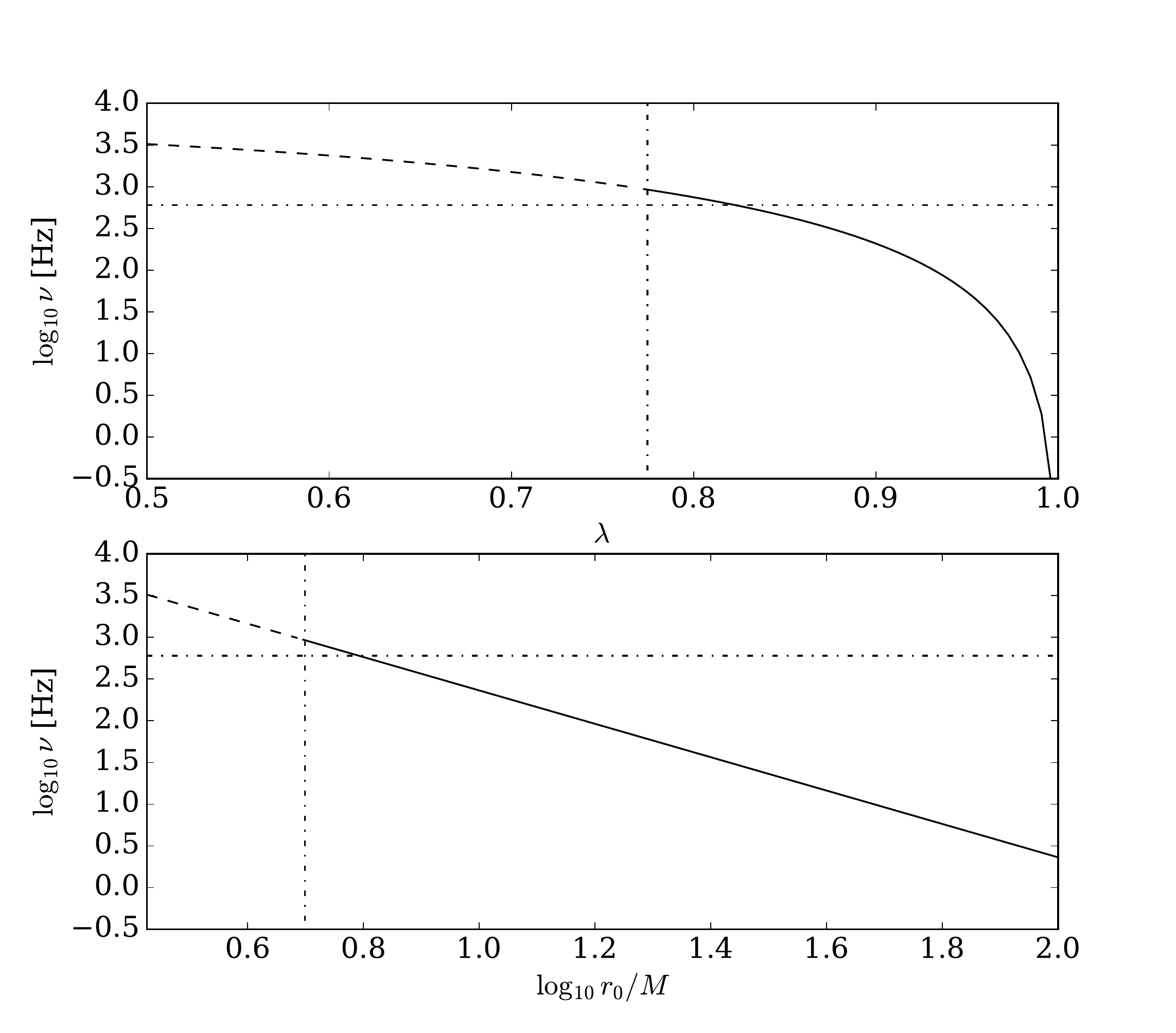}
\caption{A plot of frequency of oscillations, $\nu$, about $r_0$ with respect to the
Eddington ratio, $\lambda$, above, and with 
respect to $r_0$ below. Also shown is the interesting frequency, 600 Hz 
which corresponds to $\lambda \sim 0.82$ and the $\lambda$ for which
$r_0=5M$, below which the frequency is not physical. \label{fig:frequency}}
\end{figure}

We have calculated our angular frequency with respect to the proper time experienced by the shell, $\tau$,
so we multiply by a factor of $g_{tt}$ to redshift the angular frequency 
into the coordinate time, $t$. $\omega' = \omega \sqrt{1-2M/r_0}=\lambda \omega$. 
The frequency, $\nu=\omega'/2\pi$, is then
\begin{equation}
\nu = \dfrac{c^3}{2\pi G}\lambda\tilde\omega
\approx 8.08  (1-\lambda^2)^2
\dfrac{M_\odot}{M}\,\text{kHz}.
\end{equation}
In Fig.~\ref{fig:frequency} we can see a plot of $\nu$ as a function of $\lambda$.
An example value of $\lambda = 0.8$ with a mass of $1.4 M_\odot$ gives, $\nu=750$ Hz.

\subsection{Damped Oscillations}
At this point we calculate the strength of the radiation drag for small velocities. 
If the drag coefficient is small
then we should further explore
this and other oscillation modes as a possible explanation for X-ray burst QPOs. 
If we find that the drag coefficient is large enough to overdamp the oscillations, 
then we should
rule out this particular mode of the \textit{optically thin} atmospheres. 
Although, one should still study the
oscillatory behavior of
the \textit{optically thick} atmospheres
which may be different. 

Our equation of motion is now given by 
\begin{equation}
\dfrac{1}{M}\dfrac{d^2\xi}{d\tau^2} +\left.\dfrac{\partial \mathcal{F}}{\partial u^r}\right|_{r_0,0} \dfrac{d\xi}{d\tau}
 +  \dfrac{(1-\lambda^2)^4}{16M^3\lambda^2} \xi =0.
\end{equation}

To calculate the drag coefficient for our linearized equation of motion, we just need to evaluate the 
$u^r$ derivative of $\mathcal{F}$.

To calculate the $u^r$ derivative, we return to our equation for the flux, 
where we have already simplified the
first term. Substituting for $I(r)$ gives
\begin{equation}
F^r = \dfrac{L_\infty}{4\pi r^2} - 
 \dfrac{L_\infty}{4 \pi R^2} \dfrac{1-2M/R}{(1-2M/r)^2}
\left[\tilde{T}^{(r)(r)} 
+ \tilde{\epsilon}\right] u^r.
\end{equation}
We take the definition of $\tilde{\epsilon}$ from \cite{{stahl+13}},
\begin{equation}
\tilde{\epsilon} = \dfrac{T^{\mu \nu}u_\mu u_\nu}{\pi I(r)},
\end{equation}
 and the relevant, dimensionless tetrad components of $T^{\mu\nu}$ written as,
$\tilde{T}^{(t)(t)}$ and $\tilde{T}^{(r)(r)}$,
derived in \cite{abramowicz}, given by
\begin{align}
\tilde{T}^{(t)(t)} &=  2(1-\cos \alpha(r)),\\ 
\tilde{T}^{(r)(r)} &= \dfrac{2}{3} (1-\cos^3\alpha(r)),
\end{align}
    also in terms of the viewing angle, $\alpha(r)$. 
Substituting back into the equation of motion keeping terms to 1st order in $u^r$, we now have
\begin{equation}
\mathcal{F}(r,u^r) = \dfrac{1}{r^2}-
\dfrac{\lambda}{r^2\sqrt{1-2M/r}} + 
 \dfrac{1}{R^2} \dfrac{1-2M/R}{(1-2M/r)^2}
\left[\tilde{T}^{(r)(r)} +
\tilde{T}^{(t)(t)} \right]u^r,
\end{equation}
which makes evaluating the derivative very simple, 
\begin{equation}
\dfrac{\partial\mathcal{F}}{\partial u^r} =  
 \dfrac{\lambda}{R^2} \dfrac{1-2M/R}{(1-2M/r)^2}
\left[\tilde{T}^{(r)(r)} +
\tilde{T}^{(t)(t)} \right].
\end{equation}
At the point we only need to evaluate the derivative at $r_0$,
\begin{equation}
\left.\dfrac{\partial \mathcal{F}}{\partial u^r}\right|_{r_0,0} =  
\dfrac{1-2M/R}{\lambda^3 R^2}
\left[
\frac{2}{3} (1-\cos \alpha_0 ) \left(\cos ^2\alpha_0 +\cos \alpha_0 +4\right)
\right],
\end{equation}
where
\begin{equation}
\sin^2 \alpha(r_0) =(1-\lambda^2)^2\dfrac{\lambda^2}{4}\dfrac{(R/M)^2}{1-2M/R}.
\end{equation}

From here, it becomes much simpler to switch to dimensionless variables scaled by the neutron star mass, 
\begin{align*}
\dfrac{r}{M}\rightarrow x \\
\dfrac{R}{M}\rightarrow X \\
\dfrac{\xi}{M}\rightarrow \tilde{\xi}\\
\dfrac{\tau}{M}\rightarrow \tilde{\tau}
\end{align*}
Then the equation of motion is 
\begin{equation}
\label{eqn:dimensionlessdho}
\dfrac{d^2\tilde{\xi}}{d\tilde{\tau}^2} +M^2\left.\dfrac{\partial \mathcal{F}}{\partial u^r}\right|_{r_0,0} \dfrac{d\tilde{\xi}}{d\tilde{\tau}}
 +  \left.\dfrac{\partial \mathcal{F}}{\partial r}\right|_{r_0,0} M^3\tilde{\xi} =0.
\end{equation}

If, for convenience, we set the dimensionless quantities
\begin{align}
f &= M^3\left.\dfrac{\partial \mathcal{F}}{\partial r}\right|_{r_0,0}
=\dfrac{(1-\lambda^2)^4}{16\lambda^2},\\
g &=M^2\left.\dfrac{\partial \mathcal{F}}{\partial u^r}\right|_{r_0,0} 
=\dfrac{1-2/X}{\lambda^3 X^2}
\left[
\frac{2}{3} (1-\cos \alpha_0 ) \left(\cos ^2\alpha_0 +\cos \alpha_0 +4\right)
\right],
\end{align}
then we can classify the damping
from the sign of
\begin{equation}
\tilde{\omega}_d^2 = \dfrac{g^2}{4} - f.
\end{equation}
If $\omega_d^2 < 0$ then we have an underdamped oscillation, 
$\omega_d^2 > 0$ is overdamped and $\omega_d =0$ is critically damped. 
\begin{figure}
\includegraphics[width=\columnwidth]{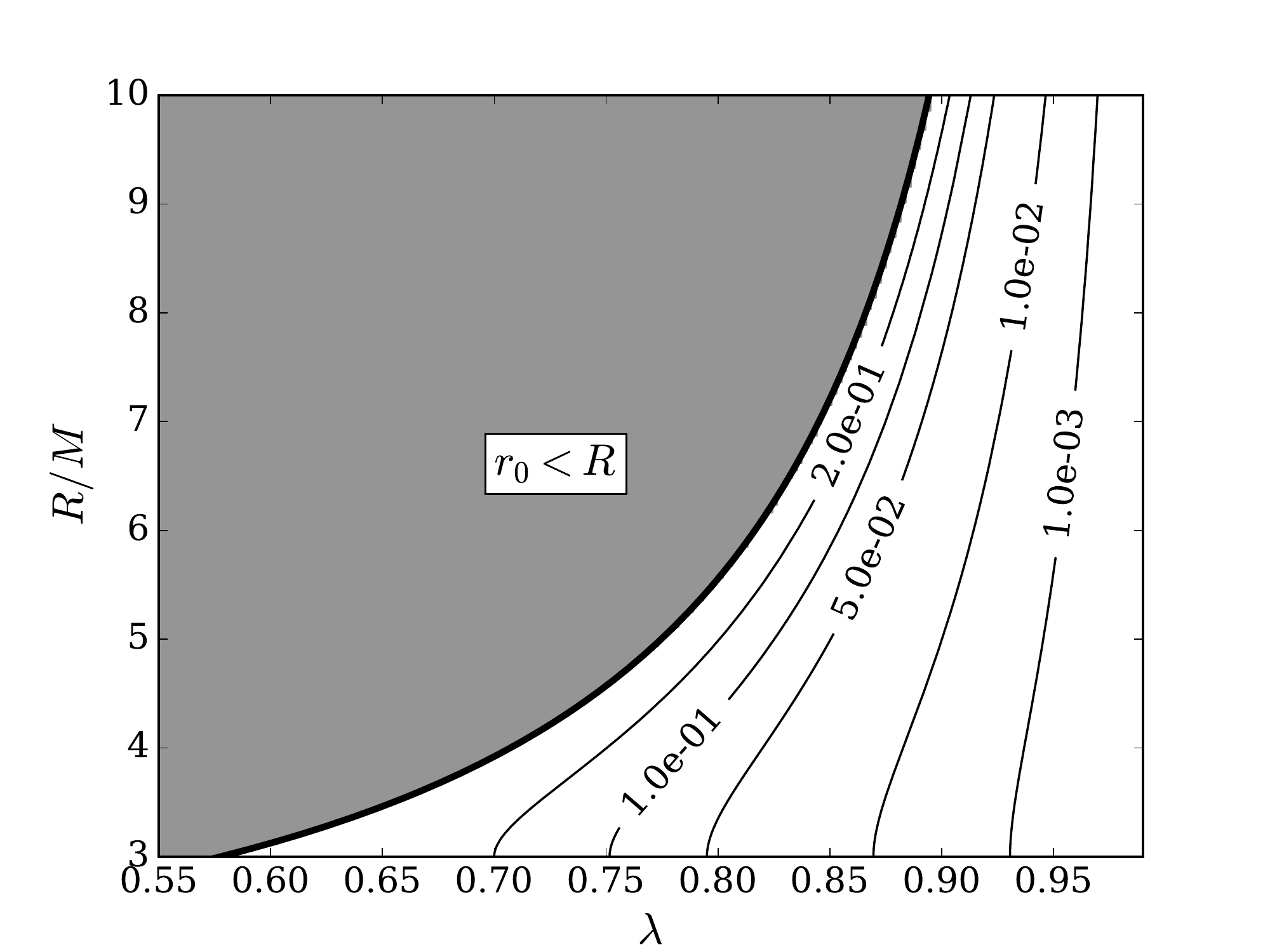}
\caption{Contours of $\tilde{\omega}_d^2/\tilde{\omega}^2$ over a reasonable parameter space
	for neutron star radius, $R$, and Eddington ratio, $\lambda$. The grey region
	on the left side of the plot corresponds to $r_0 < R$ which is unphysical. The thick line
	marks the boundary where $r_0=R$.
	There is no region on the plot where $\tilde{\omega_d}^2$ is less than zero
	which means that the thin shell eigenmode is overdamped over the parameter space
	we consider.
	\label{fig:damped}}
\end{figure}

If we take
some example values as before, $\lambda = 0.8$ and $R=5M$, then we get 
$$f \approx 0.0016, g \approx 0.089,$$
$$\omega_d \approx 0.06 = 885\,\text{s}^{-1}.$$ 

Because $\omega_d$ is real,
we have an overdamped solution with an exponential decay without oscillation.
For this reason we do not label non-oscillating quantities of dimension $[s]^{-1}$
with Hz, so that they are not to be thought of as oscillation frequencies, but rather
inverse decay timescales. In fact, over the whole reasonable parameter space of $\lambda$ and 
$R$, the eigenmode is overdamped. This can be seen in Fig.~\ref{fig:damped},
a plot of the ratio of $\tilde{\omega}_d^2$ to $\tilde{\omega}^2$. We can see
that for small $\lambda$ we are strongly overdamped and as $\lambda$ increases the 
solution approaches critical damping. 
Our damped harmonic oscillator equation yields two exponential solutions of the form 
$C_{\pm} \exp(\gamma_{\pm} \tau)$. 
where 
\begin{equation}
\gamma_\pm = -\dfrac{g}{2} \pm \sqrt{\dfrac{g^2}{4} - f}.
\end{equation}
Both of which correspond to decay constants of $\gamma_+ = -0.025, \gamma_- = -0.064 $. 
If we convert to seconds and redshift for the observer at infinity we have
\begin{equation}
\gamma_\pm' = \dfrac{c\sqrt{B(r_0)}}{2\pi r_g}\gamma_\pm= 32.31\times 10^3\,\text{s}^{-1}\,\left(\dfrac{M}{M_\odot}\right)^{-1} \lambda \gamma_\pm,
\end{equation}
$$\gamma_+'=369\,\text{s}^{-1}, \gamma_- = 945\, \text{s}^{-1}.$$
The scales of decay for each solution are then $1/\gamma_\pm$ so 
$$\tau_+ = 0.0027 \, \text{s}, \tau_- = 0.0011 \, \text{s}.$$

\section{Validity of the nonlinear regime}
A key assumption mentioned at the beginning of the work relied on the atmosphere's 
thickness and velocities being small enough to be able to linearize $\mathcal{F}(r,u^r)$. 
This allows us to obtain analytical
trajectories for the fluid for the thin shell mode, but we are interested to see the 
extent to which this linearisation is valid, both in position and velocity.
While we do not expect oscillatory 
behavior to exist at larger velocities and displacements, we can 
still study how the timescale for decay varies from that predicted by the 
analytical treatment.

To explore the degree to
which the linear regime of $\mathcal{F}$ is
valid we compare the trajectories of the thin shell mode to those of test particles 
which obey the equation of motion, 
\begin{equation}
\label{eq:testparticle}
\dfrac{d^2\tilde{\xi}}{d\tilde{\tau}^2} = - M^2 \mathcal{F}(r,u^t).
\end{equation} 
We expect the trajectories from this equation,
given by $\xi_e(t)$ 
to be similar to the trajectories given by
Eq.~\ref{eqn:dho}, which we will denote, $\xi_a(t)$. 

\begin{figure}
\includegraphics[width = \columnwidth]{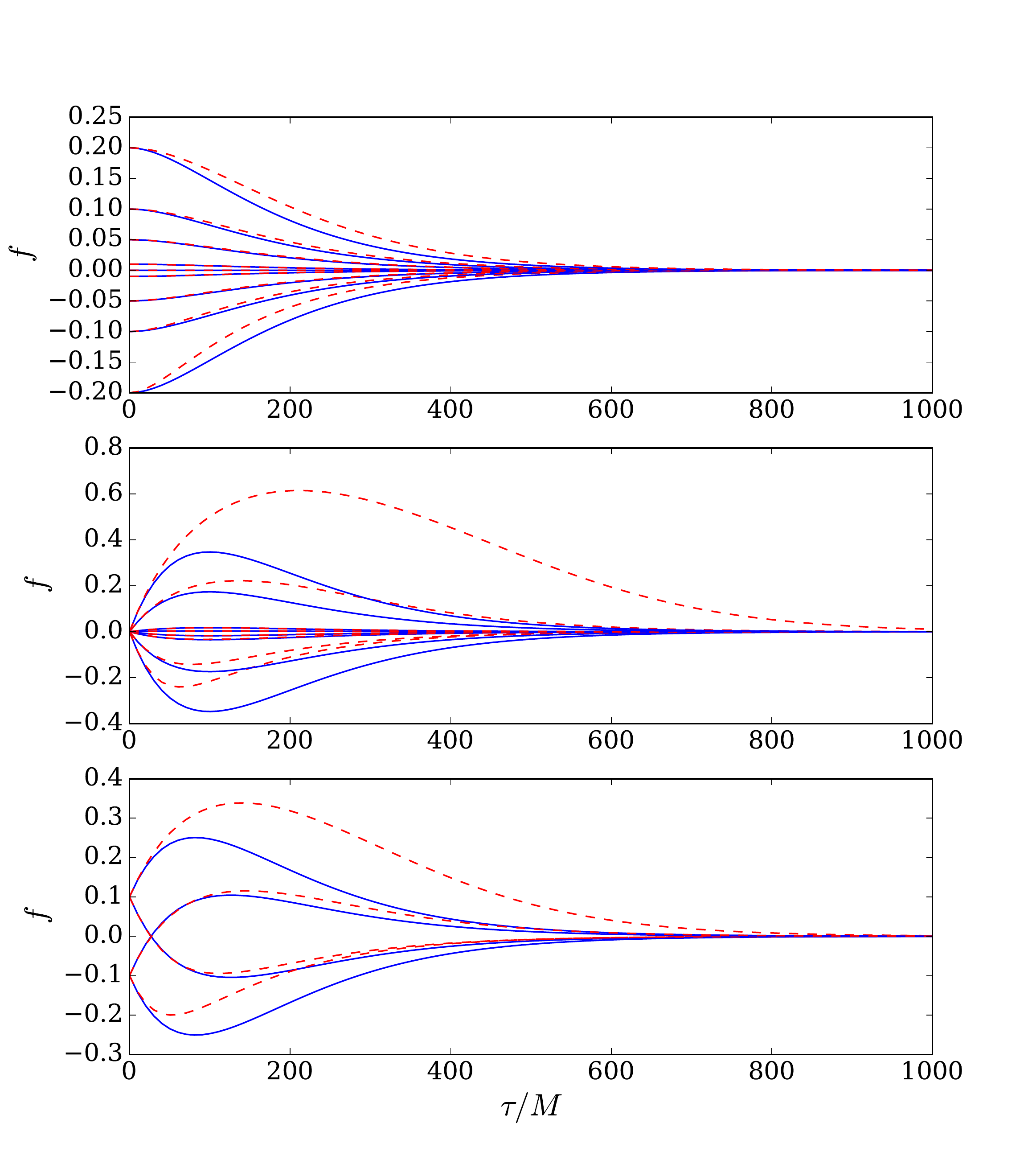}
\caption{Trajectories from a variety of initial conditions analytically 
integrated from the linearized equation of motion, Eq.~\ref{eqn:dho} (blue), and numerically 
integrated from the test particle equation of motion, Eq.~\ref{eq:testparticle} (red dashed).
 The vertical axes 
show, $f$, the fraction the initial position is from the ECS.
Upper: The initial velocity is zero and the initial position can be read from the axis.
Middle: The initial position is at the ECS and the initial velocities are given by, 
$v(0)/c = \pm 0.005, \pm 0.01, \pm 0.05, \pm 0.1$.
Lower: Initial conditions are given by 
$(f(0),v(0)/c) = (0.1, 0.05), (0.1, -0.05), (-0.1, 0.05), (-.1,-0.05).$
 \label{fig:deviations}} 
\end{figure}

To integrate this equation we use the \texttt{odeint} routine from 
the \texttt{scipy} package which relies on the \texttt{lsoda} routine from
the FORTRAN library \texttt{odepack}. 
We test a variety of initial conditions. 
We plot these against the analytical trajectories given by solving the 
linear equation of motion, of which the trajectories are just linear 
combinations of exponentials. The plots, shown in Fig.~\ref{fig:deviations},
show both trajectories from the linear equation of motion (blue) and 
from the full equation of motion (red dashed).

The first plot in Fig.~\ref{fig:deviations} shows trajectories starting with zero initial
velocity. The initial positions of the test particle are readable form the axis where
the y-axis represents the fractional displacement from the equilibrium position 
$f=(r(0)-r_0)/r_0$. We can see from the plot that deviations from the test particle solution start to occur between $f=0.1$ and $f=0.2$. The deviation 
is small however and the atmosphere settles into the equilibrium position at about
the same time as the test particle. 
We expect that for $v(0)=0$, the atmosphere trajectory 
is valid up to initial displacements of 
$10\%$ from the equilibrium solution, and a reasonable approximation up to $20\%$ away. 

The second plot in Fig.~\ref{fig:deviations} shows trajectories with an initial position
at the equilibrium position, but with different initial velocities of 
$v(0)/c = \pm 0.005, \pm 0.01, \pm 0.05, \pm 0.1$. We can immediately see that
$v(0) = \pm 0.1c$ shows vastly different trajectories, which are also highly 
asymmetrical with respect to the equilibrium position, indicating that the 
atmospheric approximation is no longer valid and even at $v(0)/c = \pm 0.5$ there are
significant deviations. We expect the linear regime in this case to hold up
to $v(0)/c = \pm 0.01$. 

The third plot in Fig.~\ref{fig:deviations} shows trajectories corresponding 
to the following initial conditions: 
$$(f(0),v(0)/c) = (0.1, 0.05), (0.1, -0.05), (-0.1, 0.05), (-.1,-0.05).$$ The two
inner trajectories show much better agreement with the test particle. This is because
their initial conditions keep them closer to the equilibrium position where the 
atmospheric
equation of motion more accurately reflects that of the test particle.

\subsection{Convergence}
\begin{figure}
\includegraphics[width = \columnwidth]{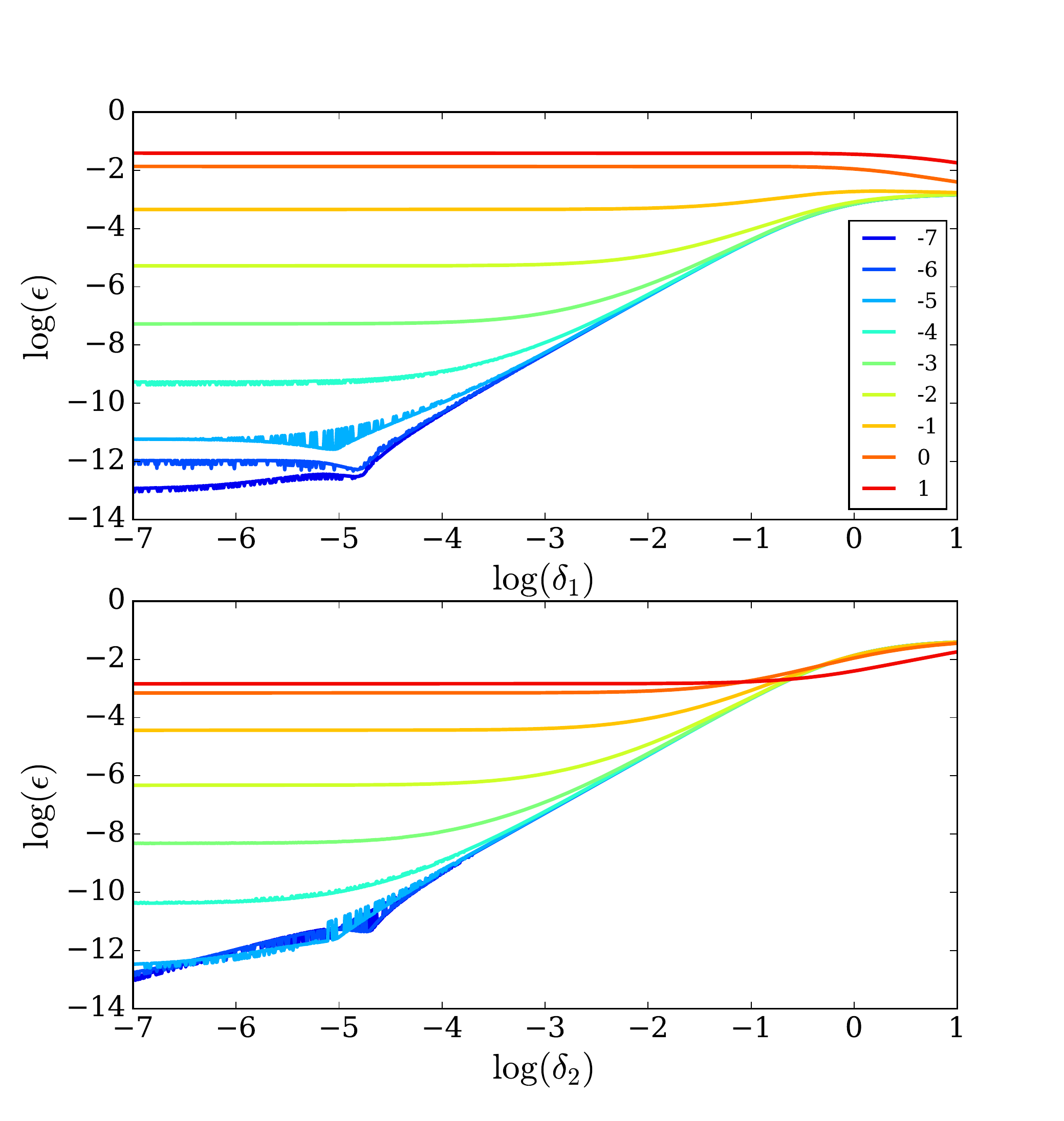}
\caption{ Here we show the results of the convergence test for the
linearisation of the equation of motion. For initial conditions
$(\xi(0), u^r(0))=(\delta_1, \delta_2 )$. The
upper panel shows different curves corresponding
to different $\delta_2$ as indicated by the legend.
The $\delta_1$s are shown on the horizontal axis. 
The lower plot shows the same except for different
curves of $\delta_1$ with $\delta_2$ on the 
horizontal axis. 
\label{fig:conversion}} 
\end{figure}

We also find it necessary to numerically confirm the convergence of the linearised fluid equation
to the test particle equation. 
In essence, we want to show that the trajectories
from the Eq.~\ref{eqn:dho} approach those from Eq.~\ref{eq:testparticle} as 
$\xi(0)\rightarrow0,u^r(0)\rightarrow 0$. We measure the deviation in the trajectory by summing
up the fractional difference along $n=100$ points, denoted by $\tau_i$,
which are evenly sampled along the trajectories shown in
Fig.~\ref{fig:deviations}. 
The error per time step is then calculated by, 
\begin{equation}
\epsilon = \dfrac{1}{n} \sum_{i=0}^n\dfrac{ \left|\xi_e(\tau_i) - \xi_a(\tau_i)\right|}{\xi_e(\tau_i)},
\end{equation}
which we calculate for different values of the initial conditions, $(\xi(0),u^r(0)) = (\delta_1, \delta_2)$.
A plot of the convergence
is shown in Fig.~\ref{fig:conversion}.
We can see that  $\epsilon$
shrinks to the level of machine precision as 
$\delta_1,\delta_2$ approach zero.

\section{Discussion and Conclusions}

\subsection{Consequences for the stability of atmospheres}
We have shown that for optically thin
levitating atmospheres, as in \cite{wielgus+thin},
 the radial,
incompressible, thin shell modes are stable against radial 
oscillations due
to the strength of the radiation drag term.
The natural frequency of these oscillations is on the order of $\sim 10^2$ Hz
if they were not overdamped. Moreover the frequency increases as the luminosity decreases.
This would lead to an increase of frequency with time if the luminosity were to decay
with time, such as during the decay phase of an X-ray burst. 
These oscillations are exactly the 
same as test particle oscillations, and the 
incompressible mode is constructed in a way that 
does not allow for extra forces due to pressure gradients. It is possible to conceive of a mode
where the pressure terms become important, such as a breathing mode where the shell
expands and contracts in opposite directions about the pressure maximum so that the 
shell becomes thinner and thicker. 
It is also possible to
construct pressure corrections to the
thin shell incompressible mode to extend its 
validity to larger thicknesses. If the eigenfrequency of such a mode is large enough, one 
can imagine that underdamped oscillations may occur, and so it would be worth exploring
such other modes.

\subsection{Optically Thick Oscillations}
In this work we have only considered optically thin solutions. \cite{wielgus+thick} have 
extended their work to include optically thick atmospheres. These atmospheres are 
constructed using numerical techniques, so it is difficult to calculate analytical 
oscillation modes.
In future work, we plan to extend this 
analysis to include numerical simulations of 
both optically thin and thick Eddington supported atmospheres. Since photons diffuse through
the optically thick atmospheres, as opposed to free streaming through the thin ones, it is
possible that the radiation drag is less efficient at damping oscillations. We except the drag to
only be effective in the optically thin edges
of the atmospheres. 

\section{acknowledgements}
The authors thank Maciek Wielgus and Omer Blaes
for useful discussions during the work.
The research was supported by the Polish
NCN grant 2013/08/A/ST9/00795.

\bibliographystyle{mn2e}
\bibliography{mybib}
\end{document}